\documentstyle[11pt,psfig]{article}
\newcommand{\AmS}{{\protect\the\textfont2
  A\kern-.1667em\lower.5ex\hbox{M}\kern-.125emS}}

\begin{document}
\title{\huge{Bottomonium Production at the Tevatron and the LHC}
\thanks{Research partially supported by CICYT under grant AEN99-0692}} 
\author{{\bf J.L. Domenech$^{a}$\thanks{domenech@evalo1.ific.uv.es} $\ $ 
and M.A. Sanchis-Lozano$^{b,c}$\thanks{Corresponding author: mas@evalo1.ific.uv.es}}
\\ \\
\it (a) Departamento de F\'{\i}sica At\'omica, Molecular y Nuclear \\
\it (b) Instituto de F\'{\i}sica
 Corpuscular (IFIC), Centro Mixto Universidad de Valencia-CSIC \\
\it (c) Departamento de F\'{\i}sica Te\'orica \\
\it Dr. Moliner 50, E-46100 Burjassot, Valencia (Spain) }
\maketitle 
\begin{abstract}
  Inclusive bottomonium hadroproduction at the Tevatron is firstly examined 
  in a Monte Carlo 
  framework with the colour-octet mechanism implemented in the event 
  generation. We extract some NRQCD colour-octet matrix elements
  relevant for $\Upsilon(1S)$ hadroproduction. 
  Remarkably we find a quite small contribution (compatible
  with zero) from feeddown of $\chi_{bJ}$ states produced
  through the colour-octet mechanism: $\Upsilon(1S)$ indirect 
  production via $\chi_{bJ}$ decays should be mainly ascribed to the
  colour-singlet model. Finally we extrapolate to LHC
  energies to predict prompt $\Upsilon(1S)$ production rates. 
\end{abstract}
\vspace{-14.5cm}
\large{
\begin{flushright}
  IFIC/99-73\\
  FTUV/99-76\\
  \today
\end{flushright} }
\vspace{14.5cm}
{\small keywords: Quarkonia production; Bottomonium; Colour-Octet Model; 
NRQCD; Tevatron; LHC}
\newpage
\section{Introduction}
B-physics has been reserved an important role in the exciting programme to be 
developed at the Large Hadron Collider or LHC along
the 21st  century. As an example, the ATLAS TDR \cite{tdr} collects a 
large number of topics related to charm and beauty flavours allowing precise
tests of the Standard Model benefitting from 
a foreseen huge statistics even with the machine running at
$\lq\lq$low" luminosity (${\simeq}\ 10^{33}$ cm$^{-2}\ s^{-1}$).

In a series of previous papers \cite{mas0,mas1,mas2,mas3}
we examined charmonium hadroproduction in a Monte Carlo framework, using
PYTHIA 5.7 \cite{pythia} event generator with the colour-octet model
(COM) \cite{braaten} implemented in.
Basically, such a production mechanism is based on the formation of
an intermediate coloured state during the hard partonic interaction, 
evolving non-perturbatively into
physical heavy resonances in the final state with certain
probabilities governed by NRQCD \cite{bodwin}. This mechanism
constitutes a (relativistic) generalization of the so-called colour-singlet
model (CSM) \cite{csm} which requires the formation of a colour-singlet state
in the hard interaction itself. Although the discrepancies 
between the CSM and experimental cross sections on bottomonia hadroproduction
are smaller than those found for charmonia \cite{fermi}, still some
extra contribution should be invoked to account for the surplus observed 
at the Fermilab Tevatron.

In this paper we extend our analysis on $J/\psi$ 
and ${\psi}'$ hadroproduction \cite{mas2} to the bottomonium family
lowest  vector resonance, i.e. the $\Upsilon(1S)$ state.
Once again, those matrix elements (MEs) determined from Tevatron
data in other analysis \cite{cho} have to be lowered once
initial-state radiation of gluons is taken into account.
This is because of the raise of the ({\em effective}) intrinsic
momentum ($k_T$) of the interacting partons enhancing the high-$p_T$ tail
of the differential cross section for heavy quarkonia production
(for more details the reader is referred to Ref. \cite{mas2}).

The study of bottomonia production 
at hadron colliders should permit a stringent test
of the colour-octet production mechanism, especially regarding
the expected (mainly transverse) polarization of the resonance created
through this process at high-$p_T$. Moreover, LHC experiments 
will cover a wider range of transverse momentum than at the Tevatron.  
Therefore, it is worth to estimate, as a 
first step, the foreseen
production rate of bottomonium resonances at the LHC and this
constitutes one of the goals of this work. Thereby any experimental
strategy  to be conducted in the future (for example a specific 
high-level trigger within the dedicated B-physics data-taking) can be 
foreseen in advance. 

We have based our study on recent results from Run 1b
at the Tevatron \cite{fermi1}. This means significantly
more statistics than the data sample from Run 1a, employed
in a former analysis \cite{cho}. However, the different sources
of prompt $\Upsilon(1S)$ production were not yet
separated along the full accessible
$p_T$-range, in contrast to charmonium production.
Hence we give in this work numerical values for some relevant 
combinations of long-distance MEs (including {\em direct} and {\em indirect} 
$\Upsilon(1S)$ production \footnote{Prompt resonance production includes 
both direct and indirect channels, the latter exclusively referred to 
feeddown from $\Upsilon(nS)$ and $\chi_{bJ}(nP)$ states - i.e. excluding 
long-lived particle decays})
extracted from the fit to the CDF experimental points. Nevertheless, we
still are able to estimate some colour-octet MEs for {\em direct} production
from the measurements on different production sources at $p_T>8$ GeV
\cite{fermi2}.

\section{Implementation of the COM in PYTHIA}

Originally the event generator PYTHIA 5.7 produces direct $J/\psi$ and
higher ${\chi}_{cJ}$ resonances via the CSM only \cite{pythia}. It is
not difficult to extend this generation to the bottomonium family
by redefining the resonance mass and wave function parameter accordingly.
In our analysis we have besides implemented a code 
in the  event  generator to account for
the colour-octet production mechanism via the following
${\alpha}_s^3$ partonic processes \footnote{We find from our simulation 
that gluon-gluon scattering actually stands 
for the dominant process as expected, gluon-quark scattering 
contributes appreciably however (${\simeq}\ 28\%$ of the colour-octet
production cross section), whereas 
quark-antiquark scattering represents ${\simeq}\ 4\%$. These fractions 
are slightly larger
than those found in charmonium production as could be expected
from a heavier quark mass} for heavy quarkonium production: 
\begin{equation}
g\ +\ g\ {\rightarrow}\ (Q\overline{Q})[^{2S+1}X_J]\ +\ g
\end{equation}
\begin{equation}
g\ +\ q\ {\rightarrow}\ (Q\overline{Q})[^{2S+1}X_J]\ +\ q
\end{equation}
\begin{equation}
q\ +\ \overline{q}\ {\rightarrow}\ (Q\overline{Q})[^{2S+1}X_J]\ +\ g
\end{equation}
where $(Q\overline{Q})[^{2S+1}X_J]$ stands for a certain heavy
quarkonium state denoted by its spectroscopic notation. In particular
we have considered  the
$^3S_1^{(8)}$, $^1S_0^{(8)}$ and $^3P_J^{(8)}$ contributions
as leading-order intermediate coloured states. In addition we generated
$\Upsilon(1S)$ and $\chi_{bJ}(nP)$ ($n=1,2$) resonances 
decaying into $\Upsilon(1S)$ according to the CSM as mentioned above. 

A lower $p_T$ cut-off was set equal to 1 GeV (by default in PYTHIA)
throughout  the generation
since some of the contributing channels are singular at vanishing
transverse momentum \cite{montp}.

\subsection{Set of $\lq\lq$fixed'' and free parameters used in the generation}

Below we list the main parameters, including 
masses and branching fractions,  used in our generation with PYTHIA 5.7.
We employed the CTEQ2L parton distribution function (PDF) in all our 
\vspace{0.2in} analysis.
\newline
{\em Masses and branching fractions:}
\begin{itemize}
\item $m_b=4.88$ GeV
\item $m_{resonance}=2m_b$
\item $BR[\Upsilon(1S){\rightarrow}\mu^+\mu^-]=2.48\ \%$ (\cite{pdg})
\end{itemize}
\vskip 0.5 cm
{\em Colour-singlet parameters} (from \cite{schuler}):
\begin{itemize}
\item $<O_1^{\Upsilon(1S)}(^3S_1)>{\mid}_{tot}=11.1$ GeV$^3$, defined as
\[ <O_1^{\Upsilon(1S)}(^3S_1)>{\mid}_{tot}\ =\ \sum_{n=1}^3
<O_1^{\Upsilon(nS)}(^3S_1)>Br[\Upsilon(nS){\rightarrow}\Upsilon(1S)X]
\]
\item $<O_1^{\chi_{b1(1P)}}(^3P_1)>=6.09$ GeV$^5$
\item $<O_1^{\chi_{b1(2P)}}(^3P_1)>=7.10$ GeV$^5$
\end{itemize}
\vskip 0.5cm
The radial wave functions at the origin (and their derivatives) 
used in the generation can
be related to the above matrix elements as
\begin{equation}
<O_1^{\Upsilon(1S)}(^3S_1)>\ =\ \frac{9}{2\pi}{\mid}R(0){\mid}^2
\end{equation}
\begin{equation}
<O_1^{\chi_{bJ(nP)}}(^3P_J)>\ =\ \frac{9}{2\pi}(2J+1) {\mid}R'(0){\mid}^2
\end{equation}
whose numerical values were obtained from
a Buchm\"{u}ller-Tye potential model tabulated in Ref. 
\vspace{0.2in} \cite{eichten}. \newline
{\em Colour-octet long-distance parameters to be extracted from the fit:}
\newline
\begin{itemize}
\item $<O_8^{\Upsilon(1S)}(^3S_1)>{\mid}_{tot}$, defined as
\begin{eqnarray}
<O_8^{\Upsilon(1S)}(^3S_1)>{\mid}_{tot} & & =\ \sum_{n=1}^3\biggl\{
<O_8^{\Upsilon(nS)}(^3S_1)>Br[\Upsilon(nS){\rightarrow}\Upsilon(1S)X]
\nonumber \\ 
& & +\ \sum_{J=0}^2
<O_8^{\chi_{bJ}(nP)}(^3S_1)>Br[\chi_{bJ}(nP){\rightarrow}\Upsilon(1S)X]
\biggr\}
\end{eqnarray}
\item $<O_8^{\Upsilon(1S)}(^1S_0)>$
\item $<O_8^{\Upsilon(1S)}(^3P_0)>$
\end{itemize}
\vskip 0.5cm
\par
On the other hand, the differences in shape between the
$^1S_0^{(8)}$ and $^3P_J^{(8)}$ contributions were not sufficiently great
to justify independent generations for them. In fact, 
temporarily setting $<O_8^{\Upsilon(1S)}(^3P_0)>
=m_b^2<O_8^{\Upsilon(1S)}(^1S_0)>$ and
defining the ratio 
\begin{equation}
r(p_T)\ =\ \frac{{\sum}_{J=0}^{2}\frac{d{\sigma}}{dp_T}[^3P_J^{(8)}]}
{\frac{d{\sigma}}{dp_T}[^1S_0^{(8)}]}
\end{equation}
it is found $r\ {\simeq}\ 5$ as a mean value over the $[0,20]$ GeV $p_T$-range.
Actually the above ratio is not steady as a function of
the $\Upsilon(1S)$ transverse momentum. Therefore in the generation we 
splitted the $p_T$ region into two domains:
for $p_T\ {\leq}\ 6$ GeV we set  $r= 6$ whereas for $p_T>6$ GeV we set
$r=4$. 

In summary, only the $^1S_0^{(8)}$ channel was generated but
rescaled by the factor $r$ to incorporate the $^3P_J^{(8)}$
contribution as we did in \cite{mas2} for charmonium hadroproduction.
Consequently, in analogy to \cite{cho} 
we shall derive a numerical estimate for the 
combination of the colour-octet matrix elements: 

\[
\frac{<O_8^{\Upsilon(1S)}(^1S_0)>}{5}+
\frac{<O_8^{\Upsilon(1S)}(^3P_0)>}{m_b^2}
\]

\subsection{Altarelli-Parisi evolution}

According to the colour-octet model, gluon fragmentation becomes the 
dominant source of heavy quarkonium direct production at high 
transverse momentum. On the other hand, 
Altarelli-Parisi (AP) evolution of the splitting gluon
into ($Q\overline{Q}$)
produces a depletion of its momentum and has to be properly taken
into account. If not so, the resulting long-distance parameter
for the  $^3S_1^{(8)}$ channel would be underestimated from the fit
\cite{montp}.

The key idea is that AP evolution of the
fragmenting gluon is performed from the evolution of
the {\em gluonic partner} of quarkonium in the final state
of the production channel 

\begin{equation}
g\ +\ g\ {\rightarrow}\ g^{\ast}({\rightarrow} 
(Q\overline{Q})[^3S_1^{(8)}])\ +\ g
\end{equation}
\vskip 0.2cm

Let us remark that, in fact,  $g^{\ast}$ is not
generated in our code \cite{mas2}.  Final hadronization into a 
($Q\overline{Q}$) bound state is taken into
account by means of the colour-octet matrix
elements multiplying the respective
short-distance cross sections \cite{cho,mas2}.
Nevertheless, it is reasonable to assume that, on the average, 
the virtual $g^{\ast}$ should evolve at high $p_T$
similarly to the other final-state gluon - which actually is
evolved by the PYTHIA machinery. 
We used this fact to simulate the (expected) evolution
of the (ungenerated)  $g^{\ast}$ whose momentum was assumed
to coincide with that of the resonance (neglecting the effect
of emission/absorption of soft gluons by the intermediate coloured 
state bleeding off colour \cite{mas3}).
\par
Therefore, event by event we get a correcting factor
to be applied to the transverse mass of the
$(Q\overline{Q})$ state (for the $^3S_1^{(8)}$ channel only):

\begin{equation}
x_p\ =\ \frac{\sqrt{P_T^{{\ast}2}+m_{(Q\overline{Q})}^2}}
{\sqrt{P_T^{2}+m_{(Q\overline{Q})}^2}} 
\end{equation}
where $P_T$ ($P_T^{\ast}$) denotes the transverse momentum of 
the final-state gluon without (with) AP evolution and
$m_{(Q\overline{Q})}$ denotes the mass of the
resonance. At high $p_T$,
\begin{equation}
 p_T^{AP}\ =\ x_p\ {\times}\ p_T
\end{equation}
where $p_T$ is the transverse momentum of the resonance
as generated by PYTHIA (i.e. without AP evolution), whereas
for $p_T\ {\leq}\ m_{(Q\overline{Q})}$ the effect becomes
much less significant as it should be. Thus the interpolation between
low and high $p_T$ is smooth with the right asymptotic
limits at both regimes.\par
The above way to implement AP evolution may appear 
somewhat simple but it remains in the spirit of our whole
analysis, i.e. using PYTHIA machinery whenever possible. In fact, 
it provides an energy depletion of the fragmenting gluon
in accordance with Cho and Leibovich's work for
charmonium hadroproduction \cite{cho,montp}.
It is worth to note, moreover, that the effect of the AP evolution
on the generation over the [0,20] GeV $p_T$-range, though sizeable, 
is considerably less pronounced for bottomonium than
for charmonium because of the larger mass of the former.\par
Notice finally that, although we can switch on/off AP evolution and
initial-state radiation {\em at will} in the generation, both next-to-leading
order effects have to be incorporated for a realistic
description of the hadronic dynamics of the process.

\vskip 1. cm
\begin{figure}[htb]
\centerline{\hbox{
\psfig{figure=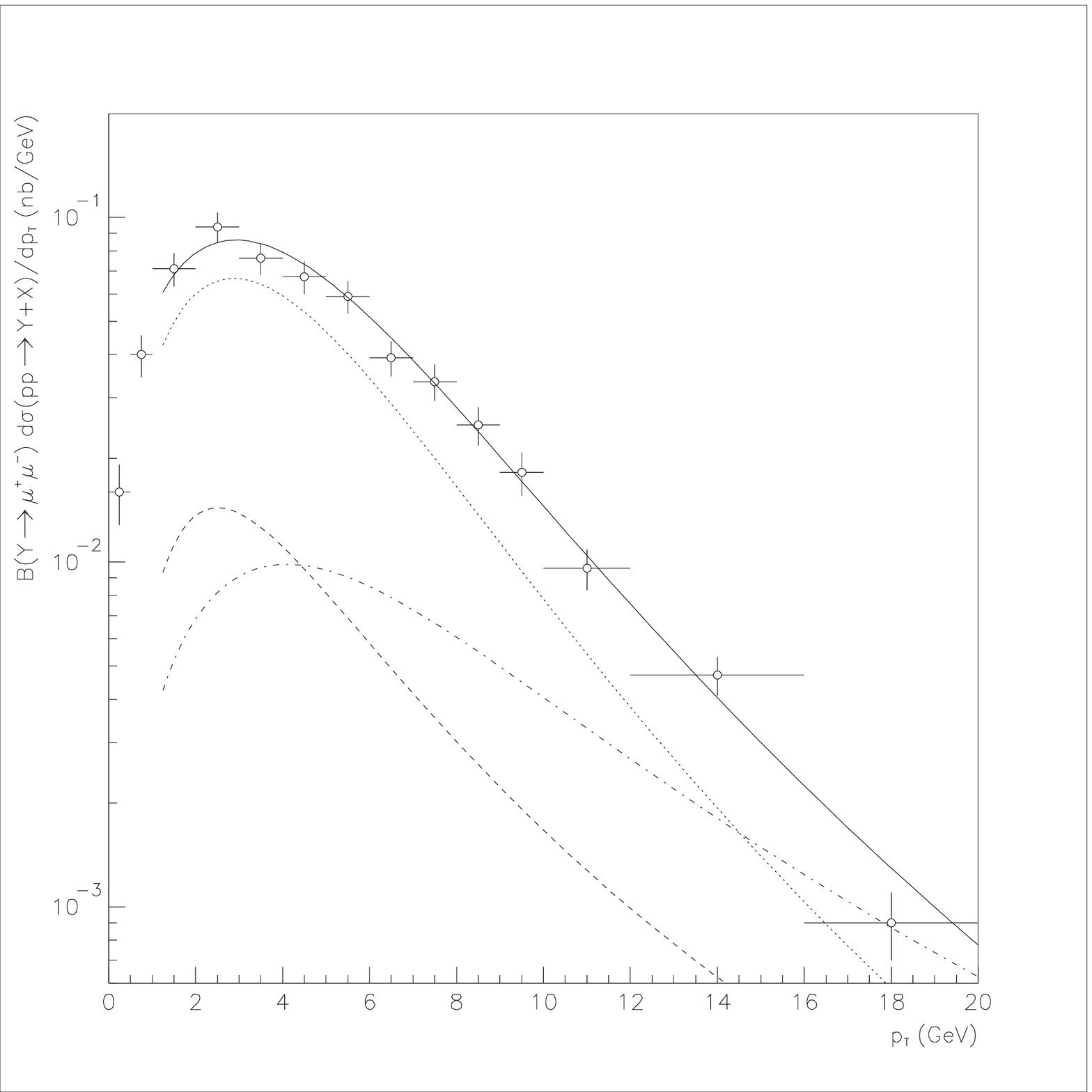,height=6.5cm,width=8.cm}
\psfig{figure=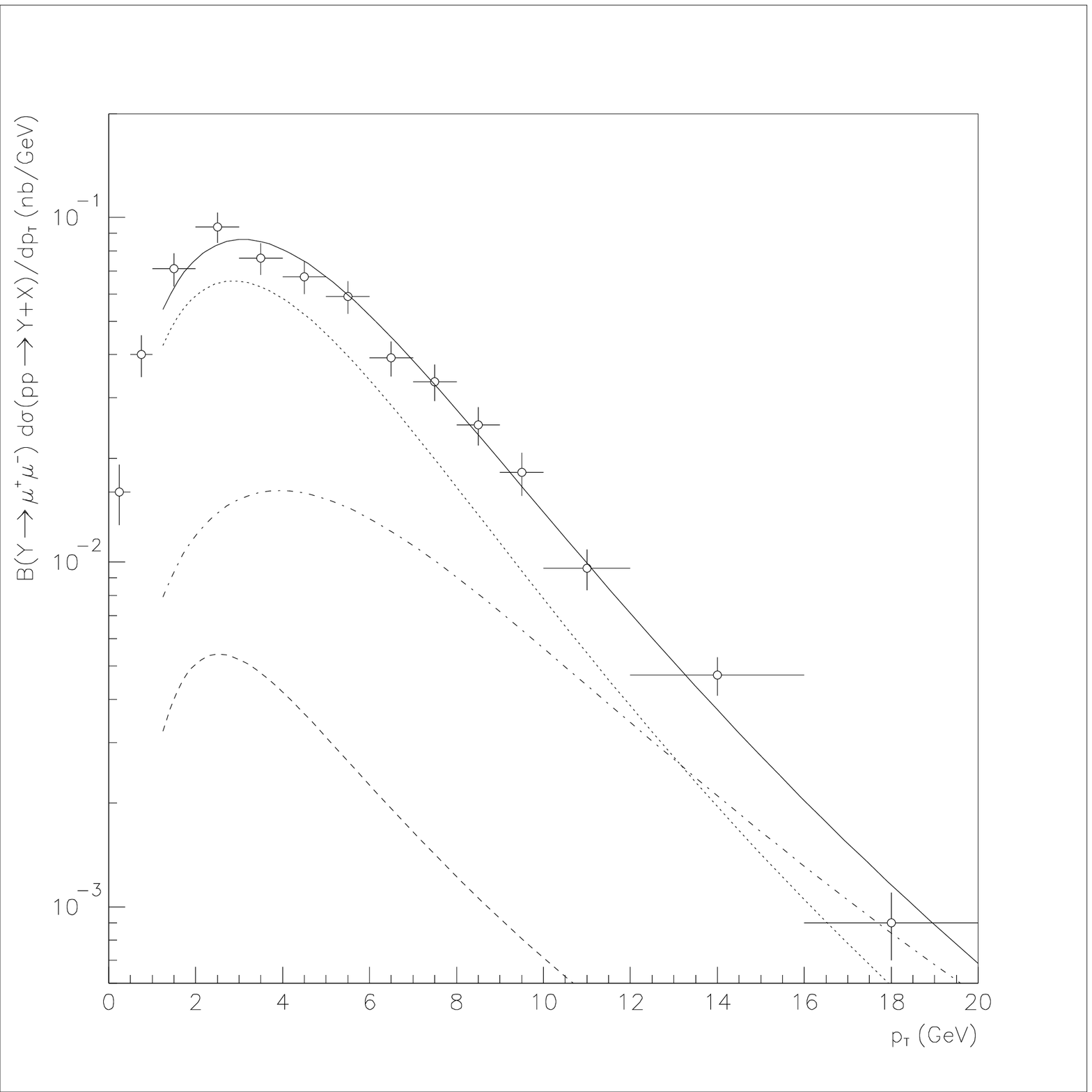,height=6.5cm,width=8.cm}
}}
\caption{Theoretical curves obtained from a fit using PYTHIA including 
the colour-octet mechanism for prompt $\Upsilon(1S)$ production against CDF
data at the Tevatron
{\it a)} without AP  evolution of the fragmenting gluon, {\it b)} 
with AP evolution of the fragmenting gluon. The  CTEQ2L parton distribution
function and $m_b=4.88$ GeV were employed in the fits; dotted line: CSM, 
dashed line: $^1S_0^{(8)}+^3P_J^{(8)}$ contribution, 
dot-dashed line: $^3S_1^{(8)}$ contribution, solid line: all contributions.} 
\end{figure}

\section{Fit to Tevatron data}

As already mentioned, the theoretical differential cross section
on inclusive production of prompt $\Upsilon(1S)$'s  stands 
above Tevatron experimental points for relatively high $p_T$ 
if the set of long-distance parameters from \cite{cho} are 
blindly employed in the PYTHIA generation with initial-state 
radiation on. Therefore we performed a new fit to recent
CDF data, incorporating both 
direct and indirect production through the
CSM (as a $\lq\lq$fixed'' contribution) which, in fact, is dominant
at low and even moderate $p_T$.
\par

\subsection{Extraction of the colour-octet  MEs}

In order to assess the effect of AP evolution on the fit parameters
we show in table 1 two sets of numerical values for the relevant
colour-octet MEs obtained from a best ${\chi}^2$
fit to Tevatron data \cite{fermi1} using the CTEQ2L PDF: (i) the first
row corresponds to a generation {\em without} AP evolution; (ii)
the second set does take it into account. Notice the
increase of $<O_8^{\Upsilon(1S)}(^3S_1)>$  in the latter case w.r.t.
AP off (but to a lesser extent than for charmonium \cite{montp}) 
whereas $M_5^{\Upsilon(1S)}$  decreases consequently to keep
the fit at low and moderate $p_T$ values.

Let us stress that our MEs numerical estimates have to be viewed with 
some caution because of the theoretical and $\lq\lq$technical'' 
(due to the Monte Carlo assumptions) uncertainties. For example our 
algorithm for AP evolution should be
regarded as a way to reasonably steepening the high-$p_T$ tail of the
(leading-order) differential cross section which otherwise
would fall off too slowly as a function of $p_T$.

\begin{table*}[hbt]
\setlength{\tabcolsep}{1.5pc}
\caption{Colour-octet matrix elements (in units of $10^{-3}$ GeV$^3$) from 
the best fit to CDF data at the Tevatron on prompt $\Upsilon(1S)$ production.
The CTEQ2L PDF was used with initial-state radiation on, and AP
evolution off and on respectively. For 
comparison we quote the values given in \cite{schuler,cho}: 
$480$ and $40{\pm}60$  
respectively.}
\label{FACTORES}

\begin{center}
\begin{tabular}{lcc}    \hline
ME:  & $<O_8^{\Upsilon(1S)}(^3S_1)>$ & 
$M_5^{\Upsilon(1S)}=5{\times}\biggl(\frac{<O_8^{\Upsilon(1S)}(^3P_0)>}{m_b^2}+
\frac{<O_8^{\Upsilon(1S)}(^1S_0)>}{5}\biggr)$ \\
\hline
AP off & $93{\pm}18$ & $17{\pm}$20 \\
\hline
AP on & $139{\pm}31$ & $6{\pm}$18  \\
\hline
\end{tabular}
\end{center}
\end{table*}
 
In figure 1 we show the theoretical curves obtained from our fit to
CDF data (independently with AP evolution off and on in the generation) 
for both colour-singlet and colour-octet contributions. Let
us remark that due to the $p_T$ cut-off parameter set in the generation, 
only those experimental points with $p_T>1$ GeV were used in the fit. 
Very good fits, with ${\chi}^2/N_{DF}$ values close to unity, were found.

\subsubsection{Separated production sources for $p_T>8$ GeV}

Current statistics does not permit
to subtract indirect production sources 
to obtain the direct $\Upsilon(1S)$ production cross section
along the full accessible $p_T$-range. Nevertheless, 
feeddown from higher states ($\Upsilon(nS)$, ${\chi}_{bJ}(nP)$)
was experimentally separated out for $p_T>8$ GeV \cite{fermi2}.
We use this information to check our analysis {\em a posteriori}
(rather than using it as a constraint in the generation)
and to draw some important conclusions. To this end
the relative fractions of the contributing  channels
for $p_T>8$ GeV are reproduced  in table 2 from Ref. \cite{fermi2}. 
On the other hand we show in table 3 
the fractions found in this work corresponding to the
different generated channels for  $p_T>8$ GeV, following the notation
introduced in section 2.1.

\vskip 0.5cm

\begin{table*}[hbt]
\setlength{\tabcolsep}{1.5pc}
\caption{Relative fractions (in $\%$) of the different contributions to 
$\Upsilon(1S)$ production from 
CDF data at $p_T>8$ GeV \cite{fermi2}. Statistical and
systematic errors have been summed quadratically.}

\label{FACTORES}

\begin{center}
\begin{tabular}{lcc}    \hline
contribution & Tevatron results \\ 
\hline
direct $\Upsilon(1S)$ & $51.8{\pm}11.4$ \\
\hline
$\Upsilon(2S)$+$\Upsilon(3S)$ & $10.7{\pm}6.4$ \\
\hline
${\chi}_b(1P)$ & $26.7{\pm}8.1$ \\
\hline
${\chi}_b(2P)$ & $10.8{\pm}4.6$ \\
\hline
\end{tabular}
\end{center}
\end{table*}

\begin{table*}[hbt]
\setlength{\tabcolsep}{1.5pc}
\caption{Relative fractions (in $\%$) of the different contributions 
to $\Upsilon(1S)$ production at the Tevatron for $p_T>8$ GeV from our 
generation. Possible contributions from
$\chi_{bJ}(3P)$ states were not generated.}

\label{FACTORES}

\begin{center}
\begin{tabular}{lcc}    \hline
contribution & our generation \\
\hline
$\Upsilon(1S){\mid}_{^3S_1^{(8)}}$ & $42.3$  \\
\hline
$\Upsilon(1S){\mid}_{^1S_0^{(8)}+^3P_J^{(8)}}$ & $3.7$  \\
\hline
$\Upsilon(1S){\mid}_{CSM}$ & $14.9$  \\
\hline
$\Upsilon(2S)$+$\Upsilon(3S){\mid}_{CSM}$ & $3.0$   \\
\hline
${\chi}_b(1P){\mid}_{CSM}$ & $21.4$   \\
\hline
${\chi}_b(2P){\mid}_{CSM}$ & $14.7$  \\
\hline
\end{tabular}
\end{center}
\end{table*}

By comparison between tables 2 and 3 we can conclude that
the $\Upsilon(1S)$ indirect production from $\chi_{bJ}$'s decays 
is almost completely accounted for by the CSM according to the
assumptions and values of the parameters presented in 
Section 2. Indeed, experimentally $37.5{\pm}9.3\%$ of 
$\Upsilon(1S)$ production is due to $\chi_{bJ}(1P)$ and
$\chi_{bJ}(2P)$ decays  \cite{fermi2}
while from our generation we find a close value, namely $36.1\%$, 
coming exclusively from colour-singlet production!   
Moreover, assuming that a $7.7\%$ from the $42.3\%$ fraction
corresponding to  the colour-octet $^3S_1^{(8)}$ contribution 
(as expressed in Eq. (6))  can be attributed to the 
$\Upsilon(2S)+\Upsilon(3S)$ channel in addition to the colour-singlet
contribution ($3\%)$, we obviously get the fraction $10.7\%$ for the latter,  
bringing our theoretical result into agreement with the
experimental value. Furthermore, this single assignment implies to 
reproduce very well the experimental fraction 
(${\approx}\ 52\%$) of direct $\Upsilon(1S)$ production by
adding the remaining $^3S_1^{(8)}$ contribution to  the
$\Upsilon(1S){\mid}_{^1S_0^{(8)}+^3P_J^{(8)}}$ and ${\Upsilon(1S)}_{CSM}$
channels (${\approx}\ 53\%$).

Of course all the above counting was based on mean values from 
table 2 and subject to uncertainties. Nevertheless, apart from
the consistency of our generation w.r.t. experimental results
under minimal assumptions, we can conclude, as an important 
consequence,  that
there is almost {\em no need for} $\Upsilon(1S)$ indirect production
from feeddown of $\chi_{bJ}$ states produced through
{\em the colour-octet mechanism}. In other words,
the relative contribution from $P$-wave states to 
$<O_8^{\Upsilon(1S)}(^3S_1)>{\mid}_{tot}$ in Eq. (6) should be 
quite smaller than
na\"{\i}vely expected from NRQCD scaling rules compared to the
charmonium sector, in agreement with 
some  remarks made in \cite{schuler}.
The underlying reason for this discrepancy  w.r.t other
analyses \cite{cho} can be traced back to
the dominant colour-singlet contribution to the cross section
at $p_T$ values as much large as $14$ GeV (see figure 1)
caused by the effective $k_T$ smearing - already applied to
charmonium hadroproduction by one of us \cite{mas2}.
 
On the other hand the corresponding velocity scaling rule
in the bottomonium sector is nicely verified. Indeed 
from the value $<O_1^{\Upsilon(1S)}(^3S_1)>{\mid}_{tot}=11.1$ GeV$^3$ and
the result found for $<O_8^{\Upsilon(1S)}(^3S_1)>{\mid}_{tot}=0.139$ GeV$^3$
shown in table 1, the ratio
\begin{equation}
\frac{<O_8^{\Upsilon(1S)}(^3S_1)>{\mid}_{tot}}
{< O_1^{\Upsilon(1S)}(^3S_1)>{\mid}_{tot}}\ {\approx}\ 0.012
\end{equation}
is in accordance  with the expected order of magnitude 
${\approx}\ v^4$, where $v$ is the relative velocity
of the bottom  quark inside bottomonium ($ v^2\ {\approx}\ 0.1$).

\begin{figure}[htb]
\centerline{\hbox{
\psfig{figure=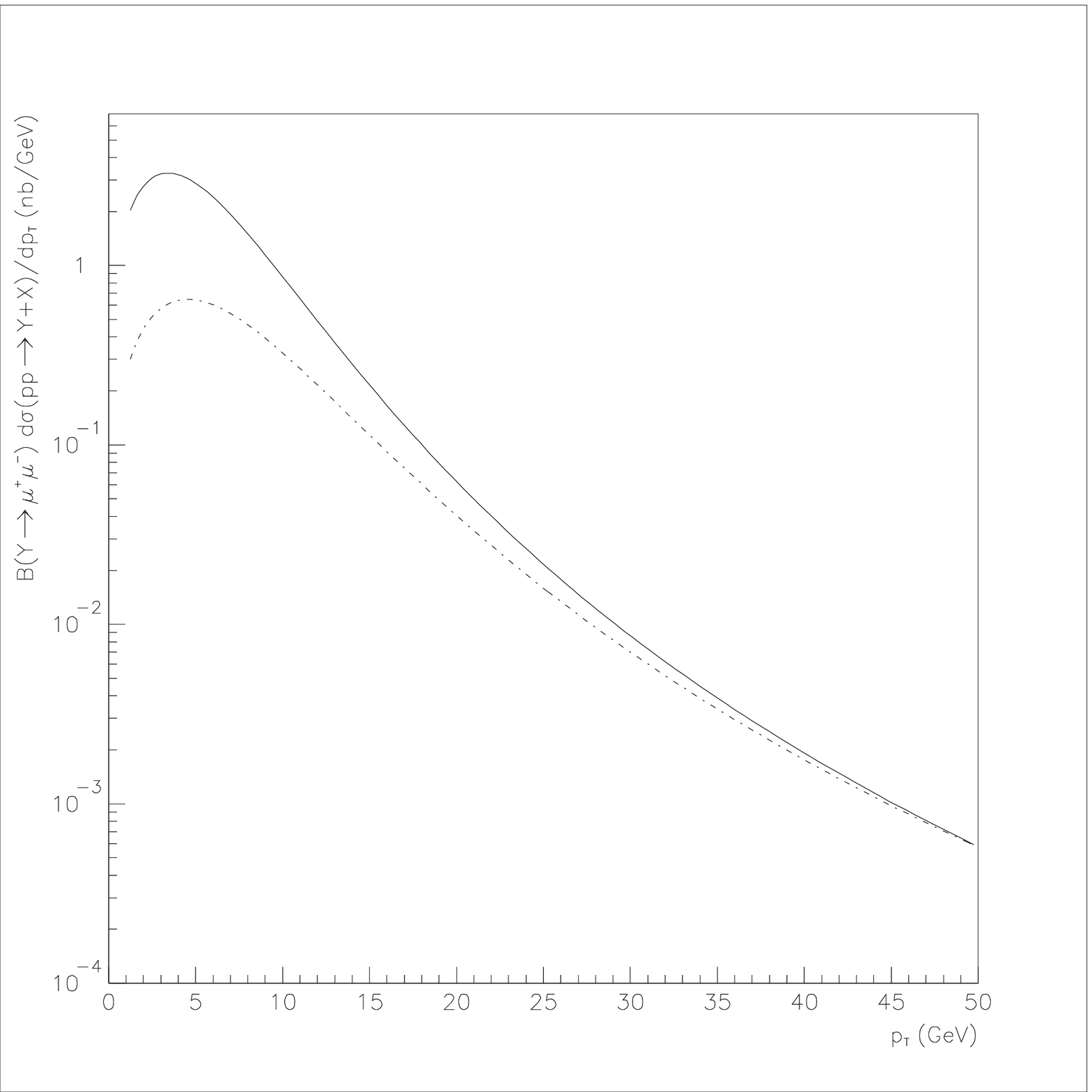,height=6.5cm,width=8.cm}
\psfig{figure=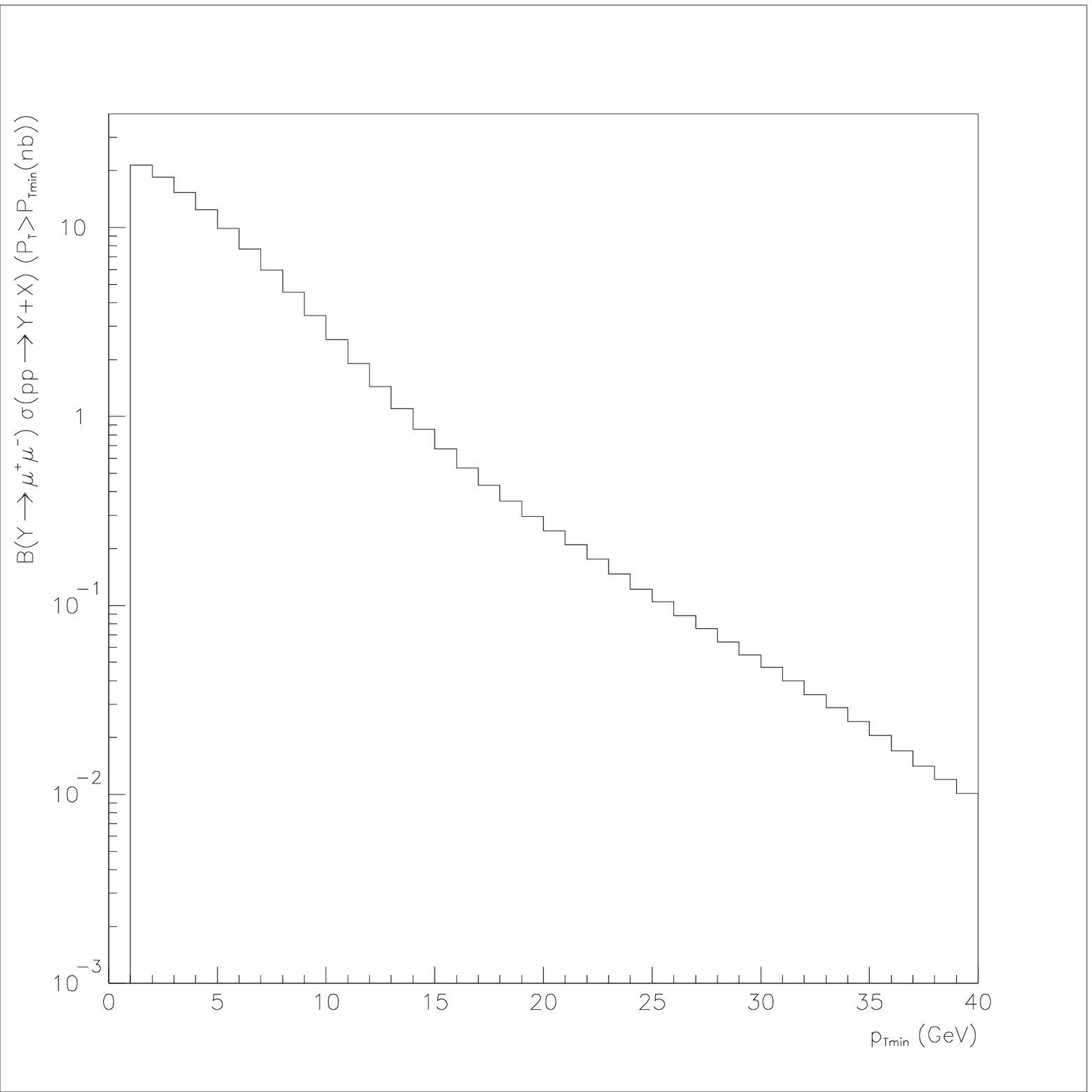,height=6.5cm,width=8.cm}
}}
\caption{{\em left :} Predicted prompt
$\Upsilon(1S)$ differential cross section at the LHC
using the CTEQ2L PDF and AP evolution incorporated in the generation.
A rapidity cut ${\mid}y{\mid}<2.5$ was required for bottomonium; 
dot-dashed line: $^3S_1^{(8)}$ contribution. Solid line: all
contributions. {\em right :} Integrated cross section.}
\end{figure}

\section{$\Upsilon(1S)$ Production at the LHC}

Bottomonium hadroproduction is especially 
interesting to check the validity of the colour-octet model as often
emphasized in the literature \cite{beneke2,tkabladze}. This becomes 
particularly clear at the LHC since experimental data will spread over
a wider $p_T$-range than at the Tevatron.\par
Keeping this interest in mind, we generated prompt 
$\Upsilon(1S)$ resonances in proton-proton collisions at 
a center-of-mass energy of 14 TeV by means of our code implemented
in PYTHIA employing the same colour-octet MEs
of table 1 with AP evolution on. We present in figure 2
our theoretical curves for the $\Upsilon(1S)$ differential and
integrated cross sections as a function of $p_T$, including both 
direct production and feeddown from higher resonance states.

\begin{figure}[htb]
\centerline{\hbox{
\psfig{figure=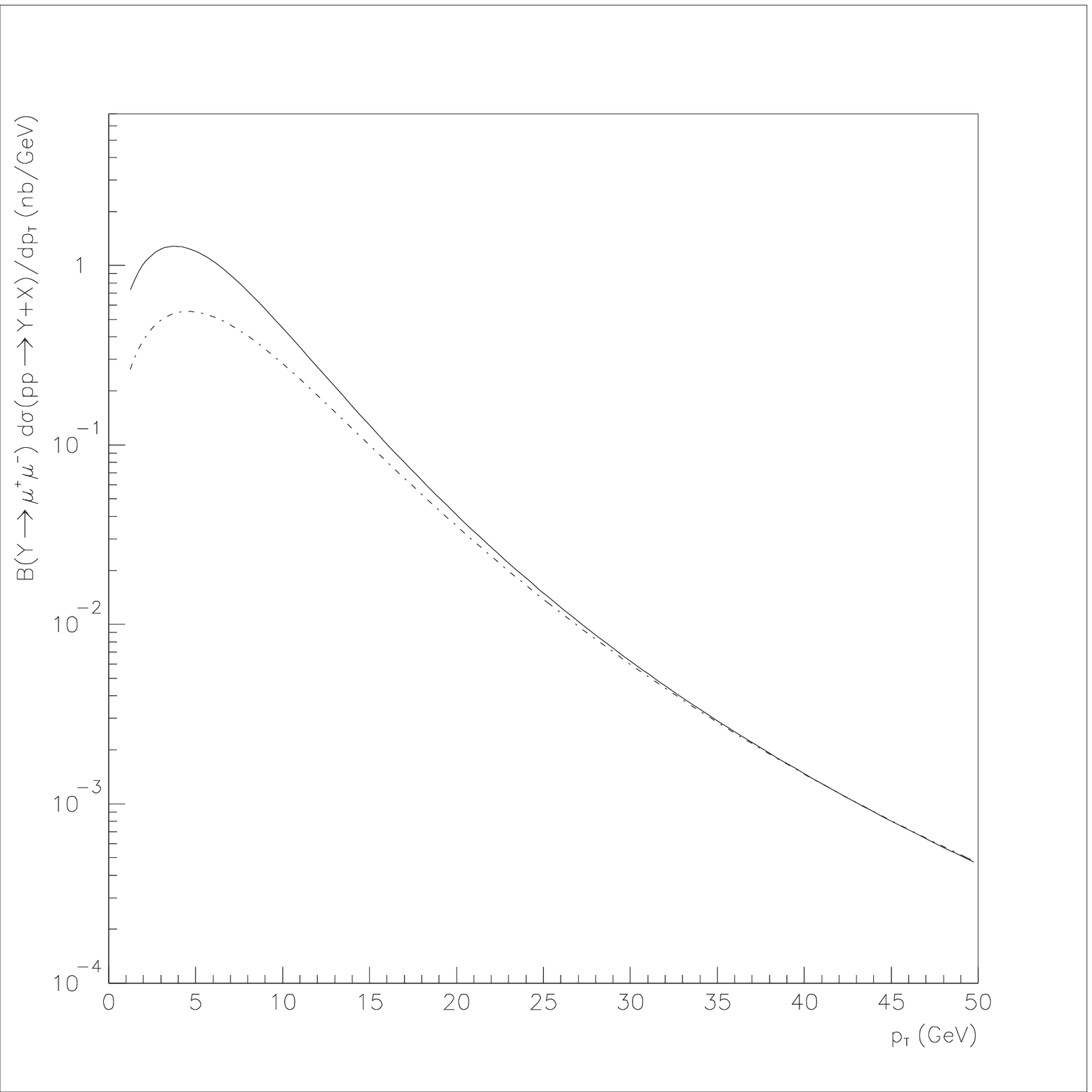,height=6.5cm,width=8.cm}
\psfig{figure=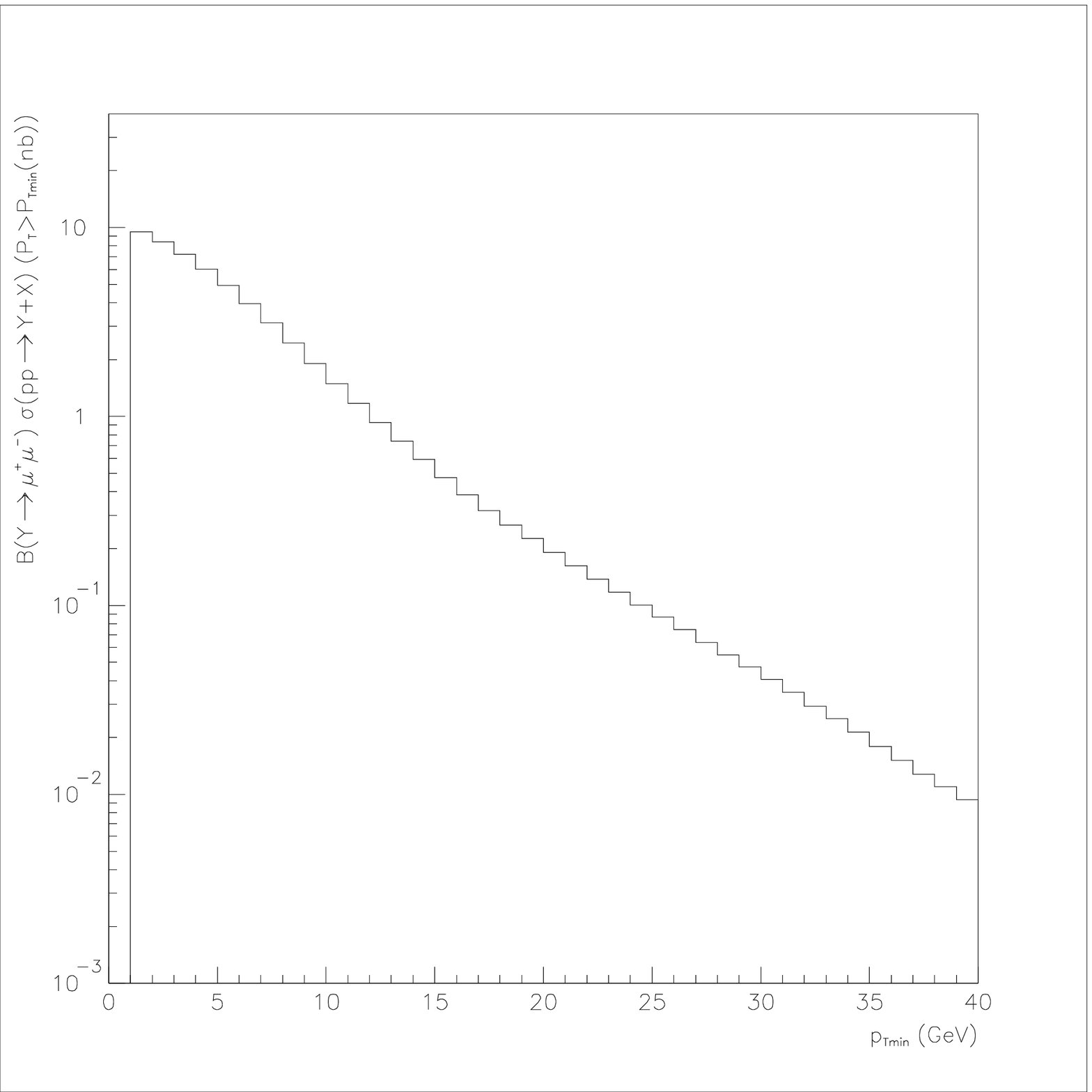,height=6.5cm,width=8.cm}
}}
\caption{The same as in figure 2 for {\em direct} $\Upsilon(1S)$
production at the LHC.}
\end{figure}

In figure 3 we show our prediction for {\em direct} $\Upsilon(1S)$
production. This
is especially interesting if LHC detectors will be able to
discriminate among such different sources of resonance
production.
To this end we generated $\Upsilon(1S)$ events
through both the CSM and COM making use of the following parameters
\begin{itemize}
\item $<O_1^{\Upsilon(1S)}(^3S_1)>{\mid}_{direct}=9.28$ GeV$^3$ 
(from \cite{schuler})
\item $<O_8^{\Upsilon(1S)}(^3S_1)>{\mid}_{direct}=0.114$ GeV$^3$
\item $M_5^{\Upsilon(1S)}=6.0$ GeV$^3$
\end{itemize}

The first value corresponds to the CSM ME for direct production. 
The $<O_8^{\Upsilon(1S)}(^3S_1)>$ ME was obtained after removing
the $\Upsilon(2S)+\Upsilon(3S)$ contribution according to
the discussion made in section 3.1.1, i.e. under the assumption that
a fraction $7.7\%$ from the $42.3\%$ in table 3 should be assigned to 
indirect production. Finally the $M_5^{\Upsilon(1S)}$ value is based on 
the assumption that this channel mainly contributes to direct production.

\vskip 1.cm

\section{Summary}

In this paper we have analyzed CDF measurements on $\Upsilon(1S)$ 
production cross sections at the Tevatron in a Monte Carlo framework. 
Higher-order QCD effects such as initial-state radiation of gluons 
and AP evolution
of splitting gluons into ($b\overline{b}$) states were taken into
account. On the other hand, since different sources 
of $\Upsilon(1S)$ production were not 
experimentally separated along the full accessible $p_T$-range
we have included all of them in the generation and later fit. Only 
for $p_T>8$ GeV feeddown from $\chi_{bJ}$ states was 
experimentally separated out from
direct production. We used such results as a consistency check
of our analysis and to draw some conclusions summarized below.

The numerical value of the
$<O_8^{\Upsilon(1S)}(^3S_1)>{\mid}_{tot}$ matrix element
should be ascribed almost totally to ${\Upsilon(nS)}$ states. 
This finding may be surprising when confronted with other
results obtained from previous analyses \cite{cho,schuler}, 
where the contribution to the ${\Upsilon(1S)}$ yield
through the colour-octet
$\chi_{bJ}$ channels was thought as dominant 
\cite{schuler,beneke,tkabladze}. 
On the contrary, we concluded from tables 2
and 3 that the {\em colour-singlet production} by itself
can account for the feeddown of $\Upsilon(1S)$ from  ${\chi}_{bJ}$
states. (Notice however that experimental uncertainties still
leave some room for a possible COM contribution but to a much lesser
extent than previously foreseen \cite{cho,schuler}.)
On the other hand the different production channels
are consistent (or can be made consistent)
with the experimental relative fractions shown in table 2, 
after some reasonable assumptions.

We have extended our study to LHC collider
experiments ($\sqrt{s}=14$ TeV center-of-mass energy).
In figure 2 we present our predictions for prompt
production rates (i.e. including direct and indirect
production) while in figure 3 we show our prediction
for direct production alone.

Lastly we conclude that the foreseen yield of $\Upsilon(1S)$'s
at LHC energy will be large enough even at
high-$p_T$ to perform a detailed analysis of the colour-octet
production mechanism and should be 
included in the B-physics programme of the LHC experiments, 
probably deserving (together with charmonia) a dedicated
data-taking \vspace{0.1in} trigger.\par

\subsection*{Acknowledgments}

We acknowledge the working subgroup on $b$-production of
the Workshop on the Standard Model (and more) at the LHC, especially
M. Kraemer and M. Mangano, for 
comments and valuable discussions. We also thank R. Cropp and
G. Feild for their assistance
on some experimental issues concerning bottomonia production
at the \vspace{0.3in} Tevatron.

\end{document}